\documentclass[preprint,showpacs,preprintnumbers,amsmath,amssymb,aps]{revtex4}

\usepackage{graphicx}
\usepackage{dcolumn}
\usepackage{bm}

\begin{document}

\preprint{BTCW-draft-Ra}

\title{Blocking temperature in magnetic nano-clusters }

\author{Burhan Bakar}
\author{L.F. Lemmens}%
 \email{lucien.lemmens@ua.ac.be}
\affiliation{%
Universiteit Antwerpen\\
Departement Fysica\\
CMI Groenenborgerlaan 171 \\
B-2020 Antwerpen \\
Belgi\"e
}%

\date{\today}

\begin{abstract}
A recent study of nonextensive phase transitions in nuclei and nuclear clusters needs  a probability model compatible with the appropriate Hamiltonian. For magnetic molecules a representation of the evolution by a Markov process achieves the required probability model that is used to study  the probability density function (PDF) of the order parameter, i.e. the magnetization. The existence of one or more modes in this PDF is an indication for the superparamagnetic transition of the cluster. This allows us to determine the factors that influence the blocking temperature, i.e. the temperature related to the change of the number of modes in the density. It turns out that for our model, rather than the evolution of the system implied by the Hamiltonian,  the high temperature density of the magnetization is the important factor for the temperature of the transition. We find that an initial probability density function with a high entropy leads to a magnetic cluster with a high blocking temperature. 

\end{abstract}

\maketitle

\section{Introduction}

Heating a ferromagnet will make it paramagnetic at a sufficiently high temperature. By cooling and magnetizing, it can be made ferromagnetic again.  This simple fact of everyday magnetism is also true on a nano-scale. As a matter of fact the existence of this phenomenon has been predicted by N\'eel \cite{LNeel} and is indicated by superparamagnetism. The temperature of the transition is the blocking temperature and  is experimentally known in molecular magnets and magnetic clusters \cite{JMMM140-379, JMMM140-1891, friedmanPRL76, PRB55-5858, PRB56-8192, TejadaPRB69}. It casts a shadow on possible applications because this temperature is very low for most materials. A theoretical estimation of this temperature could offer a better understanding of the factors that have an influence on its value  and would indicate a way to devise clusters with an appropriate  blocking temperature.
 
 As far as we know, there is no well established procedure to obtain this information, especially when the system is in the quantum regime, statistically non-extensive and the blocking temperature has to be determined from  some kind of a phase-transition in a finite system. Most theoretical approaches are based on the Fokker-Planck equation or make use of Langevin dynamics \cite{BrownJr}. 
 These methods assume that the magnetization is a continuous variable, an assumption  that is inappropriate for  systems whose magnetic degrees of freedom are small. 
 Furthermore for molecular magnets embedded in a crystalline structure the Hamiltonian $H$ describing the spectral properties of the basic molecular unit is well known \cite{GCPSSc94, gatteschiAdvMater6, gatteschiNature383, WACP01}.  
 It is used to relate the spectral properties of the system to a lot of observed responses, e.g.  quantum tunneling of magnetization \cite{gatteschiNature383, sangregorioPRL78, friedmanPRL76, GatteschiAngew}.
 Therefore a study of the blocking temperature should take the information obtained from the spectrum into account.

 Recently some substantial progress was made in non-extensive statistical physics for nuclear clusters \cite{Gross1}. 
 In the same line of thought and more important for the approach we will use here, is the observation by Gulminelli \cite{gulminelliPRE64}  
 about the equivalence of the location of the zeroes of the partition function in the complex temperature plane and the modes of the probability density function (PDF) of the order parameter. Investigations along this line include: metallic clusters \cite{Dauxois}, pairing in nuclei \cite{MelbyPRL83, SchillerPRC63}, and clusters with a negative heat capacity \cite{SchmidtPRL86, GobetPRL89}. It was first shown that for non-extensive systems the location of Yang Lee zeroes reveals the transition from  one phase to another \cite{grossmann_ZPhys207, bormannPRL84, StamerjohannsPRL88}.  
 Subsequently the equivalence between the transition  and the modality of the PDF of the order parameter offers a very direct way to obtain the temperature range where one may expect a change from ferromagnetic to paramagnetic behavior in the case of the nano-magnets. Indeed ferromagnetic behavior corresponds with a two-mode PDF for the magnetization, paramagnetic behavior with a one-mode PDF.  

 In order to calculate the PDF of the magnetization for a molecular magnet or cluster  we need a probability model able to describe cooling and heating  and compatible with that  Hamiltonian. Our starting point is a generalized propagator of the form: 
 \begin{equation}
K(m,m',\beta)=\langle m \mid \exp (-\beta H)\mid m'\rangle \frac{g_{m}}{g_{m'}}
\label{eq:propdef}
. \end{equation}
In some cases \cite{LNAW, LLPLA96}, the appropriate choice of the states $m$ and the function $g_{m}$ will lead to an evolution equation for the transition probability of a Markov chain. For the given $H$ we could indeed map the Hamiltonian description to a Markov process in a space spanned by the possible outcomes of the magnetization of the cluster. The process, only defined for $\beta>0$, describes the cooling of the system. In order to obtain the inverse evolution, the Bayes theorem can be invoked to define the adjoined heating process. In our  example the dimensionality of the process remains low. Therefore only standard  numerical methods  are necessary to perform the calculations.

Our task will be to obtain this PDF for a system with an evolution described by the process generated by the appropriate Hamiltonian. Once this calculation is done, it is our hope to find a functional relation between the parameters in the Hamiltonian and the blocking temperature. It turns out that other factors play a more important role. In order to identify these factors we will introduce in the second section, the Hamiltonian, the mapping to a Markov process, how to describe cooling and heating and outline the numerical techniques to obtain the PDF of the magnetization. In the third section we present a study of the PDF of the magnetization. We point out, by means  of several cases, that preparation, exactly as with ordinary magnets, is of fundamental importance. We summarize our results. In the last section we discuss the method, the results and try to give some perspectives.  

\section{Construction of a probability model}
The  model Hamiltonian for a molecular magnet takes the standard form \cite{friedmanPRL76, gatteschiNature383, sangregorioPRL78, caciuffoPRL81, wernsdorferPRL82, garaninPRB56}:
\begin{equation}
H=D({S}^{z})^2 +E(({S}^{x})^2-({S}^{y})^2)  +g\mu_{B}( {b_{z}}{S^{z}}+{b_{x}}{S^{x}}). \label{eq:GSH}
\end{equation}
where $D$ and $E$ are the diagonal and non-diagonal anisotropy constants, respectively and $b_{x}(b_{z})$ is the external field applied along the specified axis. 
Rewriting (\ref{eq:GSH}) and replacing the cartesian spin components by raising and lowering operators, one can split the Hamiltonian into three parts: 
\begin{equation}
H=H_{0}+V_{1}+V_{2}   \label{HBKmf}
\end{equation}
The term $H_{0}$ is diagonal in the z-component of the spin.
The term $V_{1}$ of the Hamiltonian is proportional to the $x$-component of the magnetic field and linear in the lowering and raising operators: 
\begin{equation}
V_{1}=  \frac{g\mu_{B}b_{x}}{2}\left( S^{+}{+}S^{-}\right). \label{HBKV1}
\end{equation}
The term $V_{2}$ is proportional to $E$ and quadratic in the lowering and raising operators:
\begin{equation}
V_{2}=  \frac{E}{4}\left( S^{+}S^{+}{+}S^{-}S^{-}\right). \label{HBKV2}
\end{equation} For some molecular magnets this term is forbidden by symmetry and a fourth order term is introduced. Our model allows that a term  containing a product of a lowering and a raising operator can be rewritten in terms of the z-component. 

\subsection{The evolution equation}

In order to derive the evolution equation let us consider first  $ V_{1}$.
The states are determined by two numbers $N$ and $m$: in terms of a spin $\frac{1}{2}$  system $N$ represents the number of spins, the number $m$ is
an integer of the interval $\left[ 0,N \right] $ and represents the number of spins with their magnetic component in the opposite direction. In general the magnetic quantum number is given by:
\begin{equation}
S^{z}\left| m\right\rangle  =\left( m-\frac{N}{2}\right) \left|
m\right\rangle .
\label{eq:Szm}
\end{equation}
We have chosen the number $m$ to represent the states because it allows a direct way to monitor the transitions by the counting the number of spins in the opposite direction. Lowering or raising the  number $m$ is achieved by
 the other spin operators  that  act as follows on a state: 
\begin{eqnarray}
S^{+}\left| m\right\rangle  &=&\sqrt{\left( N-m\right)
\left( m+1\right) }\left| m+1\right\rangle , \\
S^{-}\left|m\right\rangle  &=&\sqrt{\left( N-m+1\right) m}%
\left| m-1\right\rangle .
\end{eqnarray}
The matrix elements of the evolution equation in terms of these states are readily obtained from:
\begin{equation*}
\left\langle k\right| V_{1}\left| m\right\rangle = \frac{g\mu_{B}b_{x}}{2}\left( \sqrt{\left( N-m\right) \left( m+1\right) }\delta
_{k\ m+1}+\sqrt{\left( N-m+1\right) l}\delta _{k\ m-1}\right) . 
\end{equation*}
The square roots in this representation arise through the normalization of
the states necessary for the Hermitian character of the Hamiltonian and they
are eliminated by the choice $g_{m}=\sqrt{\left( 2s-m\right) !\left(
m\right) !}$ in (\ref{eq:propdef}) meaning that the normalization factor is transferred to the ket
of the quantum state. 

The matrix elements of $V_{2}$ are straightforward to calculate using the same technique and with the same $g_{m}$ the propagator becomes
\begin{equation}
\label{eq:prop }
K(m,m',\tau)=\langle m \mid \exp (-\tau H)\mid m'\rangle \bigg\{\frac{m!(N-m)!}{m'!(N-m')!}\bigg\}^{\frac{1}{2}}.
\end{equation}
and  the evolution equation of the model reading now:
 \begin{equation}
\label{eq:evoK}
\partial_{\beta}K(m,m',\beta)=\sum_{\delta=\pm1\,\pm2 } w(m,m+\delta) K(m+\delta,m',\beta) -\epsilon_{m} K(m,m',\beta), 
\end{equation} 
 does not contain square roots anymore. The energy spectrum of the diagonal part of the Hamiltonian (\ref{eq:GSH}) is given by
 \begin{equation}
\epsilon_{m}=D(\frac{N-2m}{2})^{2}-g\mu_{B} b_{z}\frac{N-2m}{2}
\end{equation}
 and the rates $w(m,m+\delta)$ are derived from the  term $V_{1}+V_{2}$.
 These rates are:
\begin{eqnarray}
w(m,m+1) & = & b(N-m) \\ \nonumber
w(m,m-1) & = & b m \\ \nonumber
w(m,m+2) & = & E(N-m)(N-m-1) \\ \nonumber
w(m,m-2) & = & E m(m-1) 
\end{eqnarray}
where $b=g\mu_{B} b_{x}$ and $E $ are chosen to be  positive. For our model (\ref{eq:GSH}) it can be done without loss of generality.  
It should be noted that the transitions do not lift the degeneracy of a configuration with magnetization characterized by $m-\frac{N}{2}$. The evolution equation (\ref{eq:evoK}) can be written as a matrix equation with dimension $(N+1)\times (N+1)$. The formal solution of this equation is:
\begin{equation}
\label{eq:solK}
\mathbf{K}(\beta)=\exp(-\beta(\mathbf{E}-\mathbf{W})),
\end{equation} 
where $\mathbf{E}$ is a diagonal matrix with $\epsilon_{m}$ on the diagonal. The matrix $\mathbf{W}$ has the rates $w(m,l)$ as elements.

Although for modest values of $N$ the matrix (\ref{eq:solK}) can be calculated directly using symbolic algebra or numerically using standard techniques, the calculation has to  fit in the probability theoretical setting mentioned in the introduction. This can be done by calculating the energy and multiplicity of the quantum states and use the Gibbs rule to relate these energy eigenvalues to probabilities. This road has been followed in the references \cite{PRE82_1402, bormannPRL84, gulminelliPRE64, PRE64_047105, PRE66_046108, PhysicaA305}.
The alternative is to introduce a Markov representation for the quantum evolution (in imaginary time or equivalently in inverse temperature) of the system.

\subsection{The Markov representation}
Before we illustrate the method used to transform the system described by (\ref{eq:evoK}), we will argue on the opportunity to choose a Markov chain for the evolution. First of all the form of the Hamiltonian allows this kind of representation, as far as we know the mapping is limited by constraints on the transition rates and therefore not general. Furthermore once a mapping can be found it is straightforward to construct a probability model without further assumptions. In other words the mapping  makes  the choice of eigenstates of the magnetization as the state space  optimal with respect to the evolution inducing  transitions in the state space. The Markov property assures thereby that transition probability to go to the next state is independent of the history preceding the transition into the actual state.

The evolution equation can be split into two parts:
\begin{equation}
\partial_{\tau}K(m,m',\tau)=\sum_{\delta=\pm1\,\pm2 } w(m,m+\delta) K(m+\delta,m',\tau) -\nu_{m} K(m,m',\tau)-(\epsilon_{m}-\nu_{m} ) K(m,m',\tau),
\end{equation}
where the following part describes a Markov chain 
\begin{equation}
\partial_{\tau}K^0(m,m',\tau)=\sum_{\delta=\pm1\,\pm2 } w(m,m+\delta) K^0(m+\delta,m',\tau) -\nu_{m} K^0(m,m',\tau) 
\end{equation}
provided $\nu_{m}$ is given by:
\begin{equation}
\nu_{m}= \sum_{\delta=\pm1\,\pm2 } w(m,m+\delta).
\end{equation}
On the condition that $(\epsilon_{m}-\nu_{m} )>0$ it is possible to use $K(m,m',\tau)$ to define a probability model for the evolution of the system. 
Under this condition it is straightforward to show that
$K(m,m',\tau)>0$. A heuristic proof is given in the appendix. However in order to fall in the category of Markov chains it is necessary that $  \sum_{states}K(m,state,\tau)=1$. 
For $K^0(m,m',\tau)$ one can easily prove this property. 

For $K(m,m',\tau)$ one can construct an equivalent Markov chain by adding an extra state $c$ to the state space. Using this state which is an absorption state sometimes called the coffin-state \cite{williams}, 
one obtains:
\begin{equation}
\sum_{m'}K(m,m',\tau)+K(m,c,\tau)=1
\end{equation} 
and defining $K(c,m',\tau)=0$ for all $m'\neq c $ and $K(c,c,\tau)=1$ one obtains the new process.

This means that we obtained  a probability model with the conditional probability given by:
\begin{equation}
\label{eq:trans}
P[M_{\tau}=m\mid M_{\tau'}=m']=K(m',m,\tau-\tau'),
\end{equation}
provided $\tau>\tau'$. $M_{\tau}$ is a random variable that takes $[m=0..N]$ or $c$ as a value.  The conditional probability is derived from 
the solution of the evolution equation. In terms of the inverse temperature it should be noted that the evolution is only represented by a Markov process when the increments are positive. 
\subsection{Cooling and heating}

Cooling is described in a straightforward way: noting that in a model that describes the evolution to equilibrium the variable $\tau$ represents the inverse temperature, a solution of the evolution equation allows to calculate the conditional probability to reach a given state at $\tau$ given the state of the system at $\tau'$. Because $\tau>\tau'$ one deals with cooling. 

In order to describe heating however a further elaboration of the probability model is necessary. Assume that the inverse temperature $\beta$ is larger than $\beta'$ and suppose we know that the system is in a given state at $\beta$ then we may ask ourselves: what is the probability to find it in a state $m'$ at $\beta'$? The quantity to obtain is:
\hbox{$ P[M_{\beta'}=m'\mid M_{\beta}=m]$ }. Using Bayes' theorem one finds:
\begin{equation}
 P[M_{\beta'}=m'\mid M_{\beta}=m]=P[M_{\beta}=m\mid M_{\beta'}=m']\frac{P[M_{\beta'}=m']}{P[M_{\beta}=m]}
\end{equation}
Noting that if we know the density of the states at $\tau=0$ the probabilities $P[M_{\beta'}=m']$ and $P[M_{\beta}=m]$ can be calculated. They are given by: 
\begin{equation}
P[M_{\beta}=m]=\sum_{k}P[M_{\beta}=m\mid M_{0}=k]P[M_{0}=k],
\label{eq:Mkv}
\end{equation}
and the transition probability for heating can also be obtained from the solution of evolution equation provided the initial PDF is known.

In this section we showed that under quite general assumptions about the parameters of the model the evolution of the model under cooling or heating can be cast into a Markov chain.

\section{The PDF of the magnetization}

A key ingredient in our probability model is the density at $\beta =0$. In order to obtain the PDF of the magnetization at a given temperature either by heating or cooling the system, one has to know the initial PDF for the magnetization: $P[M_{0}=k]$. Before we analyze the temperature dependence of these initial PDF's we will comment on two possible choices. Both choices are based on the principle of equal  ``a priori'' probabilities. If the system can be described by a single spin statistically, the probability density is uniform for all states that are participating in the evolution:
\begin{equation}
\label{eq:LS}
P[M_{0}=k]=\frac{1}{N+1}.
\end{equation}
If the system allows different spin configurations leading to same magnetization, one may assume that the $M_{0}=0$ is realized by all spins in the same direction and the  $M_{0}=N$ by all spins in the opposite direction. The other states are combinations of $(N-k)$ states with spin up and $k$ states with spin down. The initial density is calculated by considering $S_{i}^{z}=\pm\frac{1}{2}$ 
spins with $S^{z}=\sum_{i}S_{i}^{z}$. The total number of spins pointing up determines the value of the total $S^{z}$ uniquely. The number of ways to realize this number is easily obtained. Taking to into account that for a spin $\frac{1}{2}$ the probability that a single spin $S_{i}^{z}$ is pointing up, is $\frac{1}{2}$, the binomial density follows:
\begin{equation}
\label{eq:MS}
P[M_{0}=k]=2^{-N}\binom{N}{ k}.
\end{equation}
These choices correspond respectively with a single large spin or with many spins all having an evolution described by the same Hamiltonian. In order to distinguish between the choice of the initial density an index will be introduced: we will use LS for an initial uniform density (\ref{eq:LS})  and MS for the binomial density (\ref{eq:MS}). 

The calculation of the PDF of the magnetization is straightforward: first the matrix elements of (\ref{eq:solK}) are calculated by a direct method. Once these matrix elements are known the transition probabilities of the Markov representation (\ref{eq:trans}) are obtained, taking the relation between  of $S^{z}$ and $M_{\beta}$ determined by their definition (\ref{eq:Szm}) into account.

\begin{figure}[htbp]
\begin{center}
\includegraphics{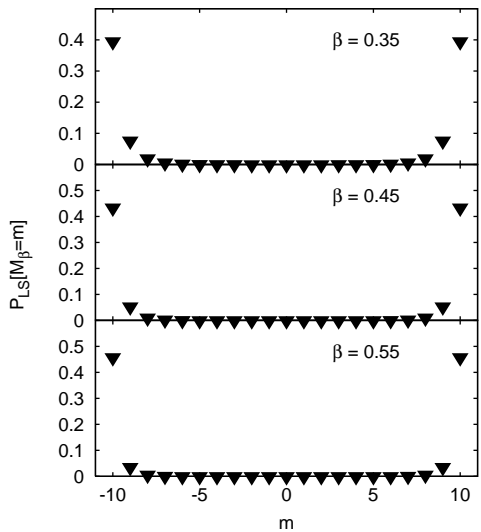}
\caption{{\bf The  probability density of the magnetization for the single spin model } The single spin model has $N+1$ realizations of the magnetization, each of these realizations is equally probable at high temperature. Using this as initial condition (\ref{eq:LS})  the density of the magnetization at a given temperature can be obtained. As illustrated here for three typical temperatures, it is seen that  the density remains bimodal in the whole temperature range.}
\label{LS-mag}
\end{center}
\end{figure}

Here we will describe the cooling and heating of the LS model first remarking that for $b_{z}=0$ and realistic values of $D$, $E$ and $b_{x}$ there is no blocking temperature i.e the PDF remains bimodal (\ref{LS-mag}). We note further that from the first and second moment of the PDF, i.e. the mean magnetization and specific heat, it is almost impossible to infer that these are moments of a bimodal density.
\begin{figure}[htbp]
\begin{center}
\includegraphics{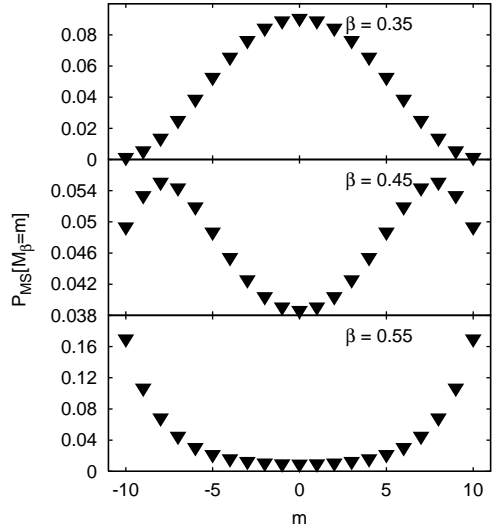}
\caption{{\bf The  probability density of the magnetization for the many spin model } The many spin model has $2^{N}$ realizations of the system leading to a binomial density of the magnetization at high temperature. Using this as initial condition (\ref{eq:MS}) the density of the magnetization at a given temperature can be obtained. As illustrated here for three typical temperatures, it is seen that the density is bimodal in the low temperature range but has a single mode in the high temperature range. The  change from a density with a single mode to a bimodal density  occurs at the blocking temperature. }
\label{MS-mag}
\end{center}
\end{figure}

Next we will describe the heating and cooling of the MS model and see that this model has a blocking temperature: changing the bimodal structure of the PDF into a PDF with one mode on heating from the low temperature phase and changing the PDF with one mode to a bimodal PDF on cooling from the high temperature phase (fig.\ref{MS-mag}). Two remarks are in order now: firstly we note that the blocking temperature has the same numerical value for cooling and for heating within the  accuracy of our calculation, secondly  the order of the transition based on the continuity of a distribution function cannot be used to classify the transition because the model has a discrete distribution function.

From the preceding analysis we arrive at a provisional conclusion: the blocking temperature of the model crucially depends on the initial density. This density is obtained by using the principle of equal \emph{a priori} probabilities for all the states participating in the evolution of the system. In the sense that we have to know the multiplicity of a state with $M_{0}=m$ in order to assign an initial PDF. 

\begin{figure}[htbp]
\begin{center}
\includegraphics{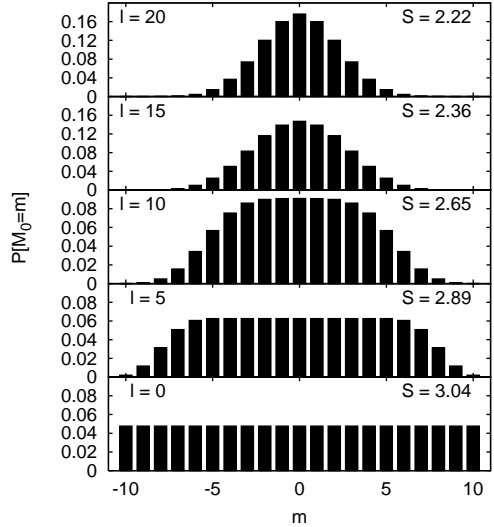}
\caption{{\bf The initial probability density of the magnetization} As the spin number $\ell$ decreases the change of the initial magnetization density from a binomial distribution $\ell=20$ to an uniform distribution $\ell=0$ (\ref{eq:ConvD}) is illustrated for $N=20$ spin cluster. The initial entropy values and $\ell$ characterizing the initial distributions are mentioned in the legend.} \label{CL-mag}
\end{center}
\end{figure}

In most molecular magnets or clusters different building blocks can be distinguished: this has led us to consider a third model where we assumed that the magnetization can be decomposed into a sum of two independent components. The first part initially has a binomial density: $$M_{0}=M_{0}^{\ell}+M_{0}^{N-\ell}$$ where $\ell$ is the number of spins. The initial density of the first part is: $$P[M_{0}^{\ell}=k]=2^{-\ell}\binom{\ell}{k},$$ the density of the second part is taken from a single spin and is uniform $$P[M_{0}^{N-\ell}=l]=1/(N-\ell+1).$$ This leads to an initial density for the magnetization given by a convolution: \begin{equation}
P[M_{0}=m]=\sum_{k \, l}\Theta(m=l+k)\frac{2^{-\ell}\binom{\ell}{k}}{(N-\ell+1)}.
\label{eq:ConvD}
\end{equation}

\begin{figure}[htbp]
\begin{center}
\includegraphics{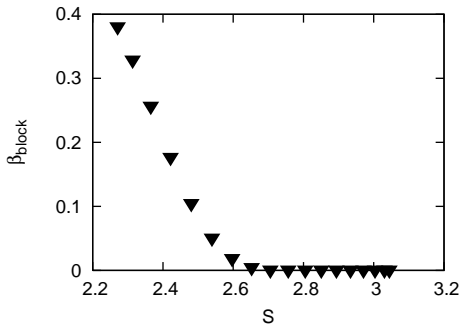}
\caption{{\bf The blocking temperature versus the initial entropy } The blocking temperature is derived from the change of the number of modes in the density of the order parameter. This change occurs at a different value of $\beta$ for a different initial density (see fig.\ref{CL-mag}). The initial density is a convolution of a uniform density with a binomial one, characterized by $S$, the entropy of the initial density.} \label{BT-mag}
\end{center}
\end{figure}

Once the initial density is chosen (fig.\ref{CL-mag}) it is straightforward to calculate the PDF of the magnetization and to determine the blocking temperature. It should be noted that the new model interpolates between the  two models previously introduced. Taking $\ell=0$ is the large spin model and taking $\ell=N$ gives the many spin model. In order to relate the blocking temperature of the model and a characteristic of the initial distribution, we have calculated the initial entropy and we have plotted the obtained blocking temperatures  (fig.\ref{BT-mag}) in function of the initial entropies.

\begin{figure}[htbp]
\begin{center}
\includegraphics{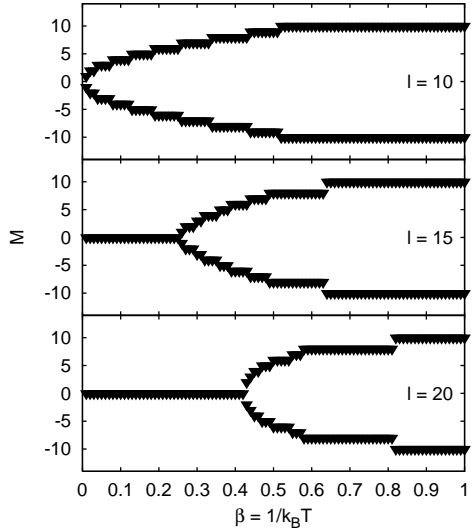}
\caption{{\bf The bifurcation of the modes for the density of the order parameter } The modes of the density of the magnetization are plotted in function of the inverse temperature $\beta$ for different initial densities (\ref{eq:ConvD}) characterized by $\ell$.} \label{BD-mag}
\end{center}
\end{figure}
The same information can be given in a bifurcation diagram (fig.\ref{BD-mag}) following the modes of the density. The blocking temperature of the preceding plot (fig.\ref{BT-mag}) is obtained by determining the value of $\beta$ where the bifurcation occurs.

\section{Discussion and conclusion}

In section 2 we mapped the evolution described by the Hamiltonian of the magnetic cluster on a continuous time Markov chain, time refers here to the ordering parameter $\beta$, the inverse temperature. This mapping is a very appealing way to obtain a probability model compatible with the Hamiltonian of the system. This model is a necessary requirement in order to use the analysis of phase transitions or finite systems mentioned in the introduction. In order  to have  a complete description, not only cooling but also heating has to be considered as a process. This is achieved by introducing an absorbing state, the so-called coffin state. Once this formalism is chosen, the procedure to calculate a PDF at a given $\beta$  requires the knowledge of an initial PDF. The reason for this requirement comes from the formulation of the evolution in terms of transition probabilities, conditional by definition.  We showed that for different choices of the initial PDF but with the same set of transition probabilities the blocking temperature critically depends on the initial condition. Our conclusion is reached by the use of a mathematically well-established procedure for Markov systems. We have to mention a possible alternative and eventually discuss the differences if any in the conclusion.

\begin{figure}[htbp]
\begin{center}
\includegraphics{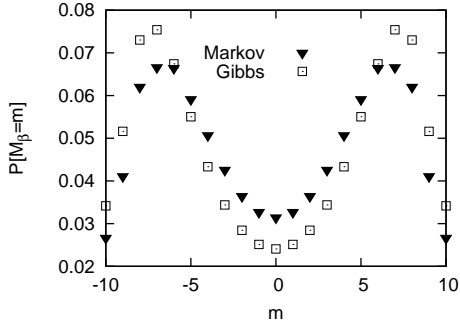}
\caption{{\bf Comparison of PDFs for Gibbs assumption and our probability model} The numerical difference between the PDF obtained in the Gibbs assumption ($\square$) (\ref{eq:Gbs})  and in our probability model($\blacktriangledown$) (\ref{eq:Mkv}) introduced in section 2 at a given inverse temperature $\beta$.}
\label{MarkovGibbs}
\end{center}
\end{figure}

The more traditional approach would be to calculate the trace of the propagator to obtain the partition function and to use the diagonal elements of the propagator explicitly in the Gibbs assumption to obtain a PDF: 
\begin{equation}
P[M_{\beta}=m]=\nu_{m}\frac{K(m,m,\beta)}{Z(\beta)}.
\label{eq:Gbs}
\end{equation}
 We have to introduce here $\nu_{m}$ \cite{BBLLUnPublish}, the multiplicity of the state $m$. In the present formalism, the initial PDF contains this information. Comparing (\ref{eq:Gbs}) with the PDF calculated in the preceding section, we find that they look very similar but there are numerical differences as can be seen in fig(\ref{MarkovGibbs}). This means  that also the blocking temperature changes slightly passing from one approach to the other. In view of the crucial factor played by the initial condition for the PDF and its relation to the multiplicity of the states, it should be noted that the initial condition is inherently related to the use of a Green function, i.e. the propagator, while in the Gibbs assumption (\ref{eq:Gbs}) it is implicit in the summation over the states. A profound analysis of the origin of the difference between (\ref{eq:Gbs}) and (\ref{eq:Mkv}) would require additional research. The main difference in approach is the choice of the states: if the eigenstates of the Hamiltonian are considered as states of the probability model there is no  reason to construct a Markov representation and the invariance of the trace with respect to cyclic permutations can be used to avoid explicitly a diagonalisation. The probability model that we considered is formulated in terms of the eigenstates of the order parameter, i.e. the eigenstates of $S^{z}$. Now $S^{z}$ and the Hamiltonian do not commute, therefore there is no common base that can serve as states of a system where both operators are diagonal. From this point of view generating  the evolution of the system in terms of the states characterised by $m$ cannot avoid  to take the transitions into account.

The model for a cluster containing for example $N$ spin $\frac{1}{ 2}$ is clearly  a many spin model, however most molecular magnets are described by a theoretical model that relies on a single large spin description. Fortunately both models obey the same dynamics. They  have the same spectrum but  a different dimensionality leading to a different blocking temperature.
In the many spin model with dimension $\ell$ the one-to-one correspondence between the magnetization and the states  implies that a specific state characterized by $m$ spins with a $s_{z}$ component  down, have a multiplicity given by $ \binom{\ell}{m}$ in order to conserve the total number of states $2^{\ell}$. As indicated  in our investigation about phase transitions in systems with  finite degrees of freedom, the multiplicity of the states is an essential ingredient. 
Compare for instance the magnetization of  the single large spin model where the spin states have  multiplicity with the many spin model with  a binomial density as initial PDF.  It is clear that the binomial multiplicity can be related to dimension. Take a harmonic oscillator in one dimension, a state with the energy level $ n \hbar \omega$ has the multiplicity $1$, while the multiplicity of that state is a function of $n$ in higher dimensions. This suggests that the single large spin model can be seen as a one-dimensional  realization of the system. According to the Van Hove theorem, it should not have a phase transition and indeed there is no transition.  However it was a surprise, for us, that the equilibrium probability density of the magnetization resembles that of the low temperature phase of the higher dimensional model, strongly indicating that one-dimensional models are low-temperature by nature.

To summarize  we do not change the evolution of the system, we change only the initial density of the magnetization by making assumptions on the magnetic structure of the cluster. The simplest magnetic structure in terms of degrees of freedom  -- each state can be realized  only in one way -- leads immediately to a bimodal density and therefore there is no transition at finite temperature. The structure with a large degree of freedom - the total number of realizations is $ 2^N$ - has a binomial initial density and leads to a finite blocking temperature. Interpolating structures with a density that arises from a convolution of  binomial densities and uniform leads to larger blocking temperatures. This means that the model indicating the smallest blocking temperature is attributed to the model with the largest number of realizations of initial  states. Less realizations of the initial states leads to larger blocking temperature. If this trend remains true for more realistic descriptions of magnetic clusters, it can be an important factor to take into account in the design of the cluster.

For a finite system with many degrees of freedom but a relatively simple evolution we could calculate the blocking temperature by mapping the system on a Markov chain in order to obtain a probability model with a change in the number of modes of the probability density function of the order parameter. This approach rests on the construction of such a probability model. It is clear that new methods are required in order to achieve such a mapping for systems described by a more realistic Hamiltonian. It should be noted that simulating the cluster belongs to the same category. In the sense that: it is relatively easy to take the diagonal elements of the propagator into account using a Metropolis algorithm \cite{BBLLNato2003}, but combining this algorithm with a Markov chain incorporating transitions in the simulation requires further investigation.

\appendix  
\section{}

In order to show that the diagonal elements of $\mathbf{K}$ can be used to define a probability, the trace should exist and the elements should be positive. The following  evolution equation is considered first:  
 \begin{equation}
\label{eq:evoKnul}
\partial_{\tau}K^{0}(m,m',\tau)=\sum_{\delta=\pm1\,\pm2 } w(m,m+\delta) K^{0}(m+\delta,m',\tau) -\sum_{\delta=\pm1\,\pm2 } w(m,m+\delta)K^{0}(m,m',\tau). 
\end{equation}
This is the evolution equation of a Markov process defined on the states $m \in [0,N]$ provided that the  $w(m,m+\delta)$ are positive. This means that with the appropriate normalization  $K^{0}(m,m',\tau)$ is a transition probability hence positive and finite. Using the transition probability $K^{0}(m,m',\tau)$ it can be shown \cite{Bremaud} that the solution of (\ref{eq:evoK})  is given by an expectation of a Feynman-Kac functional for $V(m)=\epsilon_{m}-\sum_{\delta=\pm1\,\pm2 } w(m,m+\delta)$
\begin{equation}
\label{eq:FK}
K(m,m',\tau)=\mathcal{E}_{m}[I(M(\tau)-m')\exp(-\int_{0}^{\tau}V(M(s))\!ds)]
\end{equation}
where $M(t)$ is the random process of a Markov chain with $K^{0}(m,m',\tau)$ as a transition probability and $\mathcal{E}_{m}$ is the expectation over that process. The random variable $I(M(\tau)-m')$ is an indicator necessary to satisfy the condition that the system is initially in the state $m'$. This shows that   $K(m,m',\tau)$ can be decomposed into a sum of positive terms. 
\acknowledgements
This work has been performed partly in the framework of a RAFO project/1 LEMMLU/KP99 and of the GOA BOF\ UA 2000 projects of the Universiteit Antwerpen. The authors like to thank B. Barbara and F. Brosens for discussions and suggestions.

\bibliography{BTCW.bib}

\begin{thebibliography}{37}
\expandafter\ifx\csname natexlab\endcsname\relax\def\natexlab#1{#1}\fi
\expandafter\ifx\csname bibnamefont\endcsname\relax
  \def\bibnamefont#1{#1}\fi
\expandafter\ifx\csname bibfnamefont\endcsname\relax
  \def\bibfnamefont#1{#1}\fi
\expandafter\ifx\csname citenamefont\endcsname\relax
  \def\citenamefont#1{#1}\fi
\expandafter\ifx\csname url\endcsname\relax
  \def\url#1{\texttt{#1}}\fi
\expandafter\ifx\csname urlprefix\endcsname\relax\def\urlprefix{URL }\fi
\providecommand{\bibinfo}[2]{#2}
\providecommand{\eprint}[2][]{\url{#2}}

\bibitem[{\citenamefont{N\'eel}(1949)}]{LNeel}
\bibinfo{author}{\bibfnamefont{L.}~\bibnamefont{N\'eel}},
  \bibinfo{journal}{Ann. Geophys.} \textbf{\bibinfo{volume}{5}},
  \bibinfo{pages}{99} (\bibinfo{year}{1949}).

\bibitem[{\citenamefont{Paulsen
  et~al.}(1995{\natexlab{a}})\citenamefont{Paulsen, Park, Barbara, Sessoli, and
  Caneschi}}]{JMMM140-379}
\bibinfo{author}{\bibfnamefont{C.}~\bibnamefont{Paulsen}},
  \bibinfo{author}{\bibfnamefont{J.~G.} \bibnamefont{Park}},
  \bibinfo{author}{\bibfnamefont{B.}~\bibnamefont{Barbara}},
  \bibinfo{author}{\bibfnamefont{R.}~\bibnamefont{Sessoli}}, \bibnamefont{and}
  \bibinfo{author}{\bibfnamefont{A.}~\bibnamefont{Caneschi}},
  \bibinfo{journal}{J. Magn. Magn. Mater.} \textbf{\bibinfo{volume}{140}},
  \bibinfo{pages}{379} (\bibinfo{year}{1995}{\natexlab{a}}).

\bibitem[{\citenamefont{Paulsen
  et~al.}(1995{\natexlab{b}})\citenamefont{Paulsen, Park, Barbara, Sessoli, and
  Caneschi}}]{JMMM140-1891}
\bibinfo{author}{\bibfnamefont{C.}~\bibnamefont{Paulsen}},
  \bibinfo{author}{\bibfnamefont{J.~G.} \bibnamefont{Park}},
  \bibinfo{author}{\bibfnamefont{B.}~\bibnamefont{Barbara}},
  \bibinfo{author}{\bibfnamefont{R.}~\bibnamefont{Sessoli}}, \bibnamefont{and}
  \bibinfo{author}{\bibfnamefont{A.}~\bibnamefont{Caneschi}},
  \bibinfo{journal}{J. Magn. Magn. Mater.} \textbf{\bibinfo{volume}{140}},
  \bibinfo{pages}{1891} (\bibinfo{year}{1995}{\natexlab{b}}).

\bibitem[{\citenamefont{Friedman et~al.}(1996)\citenamefont{Friedman, Sarachik,
  Tejada, and Ziolo}}]{friedmanPRL76}
\bibinfo{author}{\bibfnamefont{J.~R.} \bibnamefont{Friedman}},
  \bibinfo{author}{\bibfnamefont{M.~P.} \bibnamefont{Sarachik}},
  \bibinfo{author}{\bibfnamefont{J.}~\bibnamefont{Tejada}}, \bibnamefont{and}
  \bibinfo{author}{\bibfnamefont{R.}~\bibnamefont{Ziolo}},
  \bibinfo{journal}{Phys. Rev. Lett.} \textbf{\bibinfo{volume}{76}},
  \bibinfo{pages}{3830} (\bibinfo{year}{1996}).

\bibitem[{\citenamefont{Hernandez et~al.}(1997)\citenamefont{Hernandez, Zhang,
  Luis, Tejada, Friedman, Sarachik, and Ziola}}]{PRB55-5858}
\bibinfo{author}{\bibfnamefont{J.~M.} \bibnamefont{Hernandez}},
  \bibinfo{author}{\bibfnamefont{X.~X.} \bibnamefont{Zhang}},
  \bibinfo{author}{\bibfnamefont{F.}~\bibnamefont{Luis}},
  \bibinfo{author}{\bibfnamefont{J.}~\bibnamefont{Tejada}},
  \bibinfo{author}{\bibfnamefont{J.~R.} \bibnamefont{Friedman}},
  \bibinfo{author}{\bibfnamefont{M.~P.} \bibnamefont{Sarachik}},
  \bibnamefont{and} \bibinfo{author}{\bibfnamefont{R.}~\bibnamefont{Ziola}},
  \bibinfo{journal}{Phys. Rev. B} \textbf{\bibinfo{volume}{55}},
  \bibinfo{pages}{5858} (\bibinfo{year}{1997}).

\bibitem[{\citenamefont{Barra et~al.}(1997)\citenamefont{Barra, Gatteschi, and
  Sessoli}}]{PRB56-8192}
\bibinfo{author}{\bibfnamefont{A.~L.} \bibnamefont{Barra}},
  \bibinfo{author}{\bibfnamefont{D.}~\bibnamefont{Gatteschi}},
  \bibnamefont{and} \bibinfo{author}{\bibfnamefont{R.}~\bibnamefont{Sessoli}},
  \bibinfo{journal}{Phys. Rev. B} \textbf{\bibinfo{volume}{56}},
  \bibinfo{pages}{8192} (\bibinfo{year}{1997}).

\bibitem[{\citenamefont{Domingo et~al.}(2004)\citenamefont{Domingo, Williamson,
  G\'{o}mez-Segura, Gerbier, Ruiz-Molina, Amabilino, Veciana, and
  Tejada}}]{TejadaPRB69}
\bibinfo{author}{\bibfnamefont{N.}~\bibnamefont{Domingo}},
  \bibinfo{author}{\bibfnamefont{B.~E.} \bibnamefont{Williamson}},
  \bibinfo{author}{\bibfnamefont{J.}~\bibnamefont{G\'{o}mez-Segura}},
  \bibinfo{author}{\bibfnamefont{P.}~\bibnamefont{Gerbier}},
  \bibinfo{author}{\bibfnamefont{D.}~\bibnamefont{Ruiz-Molina}},
  \bibinfo{author}{\bibfnamefont{D.~B.} \bibnamefont{Amabilino}},
  \bibinfo{author}{\bibfnamefont{J.}~\bibnamefont{Veciana}}, \bibnamefont{and}
  \bibinfo{author}{\bibfnamefont{J.}~\bibnamefont{Tejada}},
  \bibinfo{journal}{Phys. Rev. B} \textbf{\bibinfo{volume}{69}},
  \bibinfo{pages}{052405} (\bibinfo{year}{2004}).

\bibitem[{\citenamefont{W.~F.~Brown}(1963)}]{BrownJr}
\bibinfo{author}{\bibfnamefont{J.}~\bibnamefont{W.~F.~Brown}},
  \bibinfo{journal}{Phys. Rev.} \textbf{\bibinfo{volume}{130}},
  \bibinfo{pages}{1677} (\bibinfo{year}{1963}).

\bibitem[{\citenamefont{Gatteschi et~al.}(1994)\citenamefont{Gatteschi,
  Caneschi, Pardi, and Sessoli}}]{GCPSSc94}
\bibinfo{author}{\bibfnamefont{D.}~\bibnamefont{Gatteschi}},
  \bibinfo{author}{\bibfnamefont{A.}~\bibnamefont{Caneschi}},
  \bibinfo{author}{\bibfnamefont{L.}~\bibnamefont{Pardi}}, \bibnamefont{and}
  \bibinfo{author}{\bibfnamefont{R.}~\bibnamefont{Sessoli}},
  \bibinfo{journal}{Science} \textbf{\bibinfo{volume}{265}},
  \bibinfo{pages}{1054} (\bibinfo{year}{1994}).

\bibitem[{\citenamefont{Gatteschi}(1994)}]{gatteschiAdvMater6}
\bibinfo{author}{\bibfnamefont{D.}~\bibnamefont{Gatteschi}},
  \bibinfo{journal}{Adv. Mater.} \textbf{\bibinfo{volume}{6}},
  \bibinfo{pages}{635} (\bibinfo{year}{1994}).

\bibitem[{\citenamefont{Thomas et~al.}(1996)\citenamefont{Thomas, Lionti,
  Ballou, Gatteschi, Sessoli, and Barbara}}]{gatteschiNature383}
\bibinfo{author}{\bibfnamefont{L.}~\bibnamefont{Thomas}},
  \bibinfo{author}{\bibfnamefont{F.}~\bibnamefont{Lionti}},
  \bibinfo{author}{\bibfnamefont{R.}~\bibnamefont{Ballou}},
  \bibinfo{author}{\bibfnamefont{D.}~\bibnamefont{Gatteschi}},
  \bibinfo{author}{\bibfnamefont{R.}~\bibnamefont{Sessoli}}, \bibnamefont{and}
  \bibinfo{author}{\bibfnamefont{B.}~\bibnamefont{Barbara}},
  \bibinfo{journal}{Nature} \textbf{\bibinfo{volume}{383}},
  \bibinfo{pages}{145} (\bibinfo{year}{1996}).

\bibitem[{\citenamefont{Wernsdorfer}(2001)}]{WACP01}
\bibinfo{author}{\bibfnamefont{W.}~\bibnamefont{Wernsdorfer}},
  \bibinfo{journal}{Adv. Chem. Phys.} \textbf{\bibinfo{volume}{118}},
  \bibinfo{pages}{99} (\bibinfo{year}{2001}).

\bibitem[{\citenamefont{Sangregorio et~al.}(1997)\citenamefont{Sangregorio,
  Ohm, Paulsen, Sessoli, and Gatteschi}}]{sangregorioPRL78}
\bibinfo{author}{\bibfnamefont{C.}~\bibnamefont{Sangregorio}},
  \bibinfo{author}{\bibfnamefont{T.}~\bibnamefont{Ohm}},
  \bibinfo{author}{\bibfnamefont{C.}~\bibnamefont{Paulsen}},
  \bibinfo{author}{\bibfnamefont{R.}~\bibnamefont{Sessoli}}, \bibnamefont{and}
  \bibinfo{author}{\bibfnamefont{D.}~\bibnamefont{Gatteschi}},
  \bibinfo{journal}{Phys. Rev. Lett.} \textbf{\bibinfo{volume}{78}},
  \bibinfo{pages}{4645} (\bibinfo{year}{1997}).

\bibitem[{\citenamefont{Gatteschi and Sessoli}(2003)}]{GatteschiAngew}
\bibinfo{author}{\bibfnamefont{D.}~\bibnamefont{Gatteschi}} \bibnamefont{and}
  \bibinfo{author}{\bibfnamefont{R.}~\bibnamefont{Sessoli}},
  \bibinfo{journal}{Angew. Chem. Int. Edit} \textbf{\bibinfo{volume}{42}},
  \bibinfo{pages}{268} (\bibinfo{year}{2003}).

\bibitem[{\citenamefont{Gross}(2001)}]{Gross1}
\bibinfo{author}{\bibfnamefont{D.~H.~E.} \bibnamefont{Gross}},
  \emph{\bibinfo{title}{Microcanonical Thermodynamics: Phase Transitions in
  Finite Systems}} (\bibinfo{publisher}{World Scientific, Singapore},
  \bibinfo{year}{2001}).

\bibitem[{\citenamefont{Chomaz et~al.}(2001)\citenamefont{Chomaz, Gulminelli,
  and Duflot}}]{gulminelliPRE64}
\bibinfo{author}{\bibfnamefont{P.}~\bibnamefont{Chomaz}},
  \bibinfo{author}{\bibfnamefont{F.}~\bibnamefont{Gulminelli}},
  \bibnamefont{and} \bibinfo{author}{\bibfnamefont{V.}~\bibnamefont{Duflot}},
  \bibinfo{journal}{Phys. Rev. E} \textbf{\bibinfo{volume}{64}},
  \bibinfo{pages}{046114} (\bibinfo{year}{2001}).

\bibitem[{\citenamefont{Dauxois et~al.}(2002)\citenamefont{Dauxois, Ruffo,
  Arimondo, and Wilkens}}]{Dauxois}
\bibinfo{author}{\bibfnamefont{T.}~\bibnamefont{Dauxois}},
  \bibinfo{author}{\bibfnamefont{S.}~\bibnamefont{Ruffo}},
  \bibinfo{author}{\bibfnamefont{E.}~\bibnamefont{Arimondo}}, \bibnamefont{and}
  \bibinfo{author}{\bibfnamefont{M.}~\bibnamefont{Wilkens}},
  \emph{\bibinfo{title}{Dynamics and thermodynamics of systems with long range
  interactions}} (\bibinfo{publisher}{Springer}, \bibinfo{year}{2002}).

\bibitem[{\citenamefont{Melby et~al.}(1999)\citenamefont{Melby, Bergholt,
  Guttormsen, Hjorth-Jensen, Ingebretsen, Messelt, Rekstad, Schiller, Siem, and
  {\O}deg{\aa}rd}}]{MelbyPRL83}
\bibinfo{author}{\bibfnamefont{E.}~\bibnamefont{Melby}},
  \bibinfo{author}{\bibfnamefont{L.}~\bibnamefont{Bergholt}},
  \bibinfo{author}{\bibfnamefont{M.}~\bibnamefont{Guttormsen}},
  \bibinfo{author}{\bibfnamefont{M.}~\bibnamefont{Hjorth-Jensen}},
  \bibinfo{author}{\bibfnamefont{F.}~\bibnamefont{Ingebretsen}},
  \bibinfo{author}{\bibfnamefont{S.}~\bibnamefont{Messelt}},
  \bibinfo{author}{\bibfnamefont{J.}~\bibnamefont{Rekstad}},
  \bibinfo{author}{\bibfnamefont{A.}~\bibnamefont{Schiller}},
  \bibinfo{author}{\bibfnamefont{S.}~\bibnamefont{Siem}}, \bibnamefont{and}
  \bibinfo{author}{\bibfnamefont{S.~W.} \bibnamefont{{\O}deg{\aa}rd}},
  \bibinfo{journal}{Phys. Rev. Lett.} \textbf{\bibinfo{volume}{83}},
  \bibinfo{pages}{3150} (\bibinfo{year}{1999}).

\bibitem[{\citenamefont{Schiller et~al.}(2001)\citenamefont{Schiller, Bjerve,
  Guttormsen, Hjorth-Jensen, Ingebretsen, Melby, Messelt, Rekstad, Siem, and
  {\O}deg{\aa}rd}}]{SchillerPRC63}
\bibinfo{author}{\bibfnamefont{A.}~\bibnamefont{Schiller}},
  \bibinfo{author}{\bibfnamefont{A.}~\bibnamefont{Bjerve}},
  \bibinfo{author}{\bibfnamefont{M.}~\bibnamefont{Guttormsen}},
  \bibinfo{author}{\bibfnamefont{M.}~\bibnamefont{Hjorth-Jensen}},
  \bibinfo{author}{\bibfnamefont{F.}~\bibnamefont{Ingebretsen}},
  \bibinfo{author}{\bibfnamefont{E.}~\bibnamefont{Melby}},
  \bibinfo{author}{\bibfnamefont{S.}~\bibnamefont{Messelt}},
  \bibinfo{author}{\bibfnamefont{J.}~\bibnamefont{Rekstad}},
  \bibinfo{author}{\bibfnamefont{S.}~\bibnamefont{Siem}}, \bibnamefont{and}
  \bibinfo{author}{\bibfnamefont{S.~W.} \bibnamefont{{\O}deg{\aa}rd}},
  \bibinfo{journal}{Phys. Rev. C} \textbf{\bibinfo{volume}{63}},
  \bibinfo{pages}{021306} (\bibinfo{year}{2001}).

\bibitem[{\citenamefont{Schmidt et~al.}(2001)\citenamefont{Schmidt, Kusche,
  Hippler, Donges, Kronm\"{u}ller, von Issendorff, and
  Haberland}}]{SchmidtPRL86}
\bibinfo{author}{\bibfnamefont{M.}~\bibnamefont{Schmidt}},
  \bibinfo{author}{\bibfnamefont{R.}~\bibnamefont{Kusche}},
  \bibinfo{author}{\bibfnamefont{T.}~\bibnamefont{Hippler}},
  \bibinfo{author}{\bibfnamefont{J.}~\bibnamefont{Donges}},
  \bibinfo{author}{\bibfnamefont{W.}~\bibnamefont{Kronm\"{u}ller}},
  \bibinfo{author}{\bibfnamefont{B.}~\bibnamefont{von Issendorff}},
  \bibnamefont{and}
  \bibinfo{author}{\bibfnamefont{H.}~\bibnamefont{Haberland}},
  \bibinfo{journal}{Phys. Rev. Lett.} \textbf{\bibinfo{volume}{86}},
  \bibinfo{pages}{1191} (\bibinfo{year}{2001}).

\bibitem[{\citenamefont{Gobet et~al.}(2002)\citenamefont{Gobet, Farizon,
  Farizon, Gaillard, Buchet, Carr\'{e}, Scheier, and M\"{a}rk}}]{GobetPRL89}
\bibinfo{author}{\bibfnamefont{F.}~\bibnamefont{Gobet}},
  \bibinfo{author}{\bibfnamefont{B.}~\bibnamefont{Farizon}},
  \bibinfo{author}{\bibfnamefont{M.}~\bibnamefont{Farizon}},
  \bibinfo{author}{\bibfnamefont{M.~J.} \bibnamefont{Gaillard}},
  \bibinfo{author}{\bibfnamefont{J.~P.} \bibnamefont{Buchet}},
  \bibinfo{author}{\bibfnamefont{M.}~\bibnamefont{Carr\'{e}}},
  \bibinfo{author}{\bibfnamefont{P.}~\bibnamefont{Scheier}}, \bibnamefont{and}
  \bibinfo{author}{\bibfnamefont{T.~D.} \bibnamefont{M\"{a}rk}},
  \bibinfo{journal}{Phys. Rev. Lett.} \textbf{\bibinfo{volume}{89}},
  \bibinfo{pages}{183403} (\bibinfo{year}{2002}).

\bibitem[{\citenamefont{Grossmann and Rosenhauer}(1967)}]{grossmann_ZPhys207}
\bibinfo{author}{\bibfnamefont{S.}~\bibnamefont{Grossmann}} \bibnamefont{and}
  \bibinfo{author}{\bibfnamefont{W.}~\bibnamefont{Rosenhauer}},
  \bibinfo{journal}{Z. Phys} \textbf{\bibinfo{volume}{207}},
  \bibinfo{pages}{138} (\bibinfo{year}{1967}).

\bibitem[{\citenamefont{Borrmann et~al.}(2000)\citenamefont{Borrmann,
  M\"{u}lken, and Harting}}]{bormannPRL84}
\bibinfo{author}{\bibfnamefont{P.}~\bibnamefont{Borrmann}},
  \bibinfo{author}{\bibfnamefont{O.}~\bibnamefont{M\"{u}lken}},
  \bibnamefont{and} \bibinfo{author}{\bibfnamefont{J.}~\bibnamefont{Harting}},
  \bibinfo{journal}{Phys. Rev. Lett.} \textbf{\bibinfo{volume}{84}},
  \bibinfo{pages}{3511} (\bibinfo{year}{2000}).

\bibitem[{\citenamefont{Stamerjohanns et~al.}(2002)\citenamefont{Stamerjohanns,
  M\"{u}lken, and Borrmann}}]{StamerjohannsPRL88}
\bibinfo{author}{\bibfnamefont{H.}~\bibnamefont{Stamerjohanns}},
  \bibinfo{author}{\bibfnamefont{O.}~\bibnamefont{M\"{u}lken}},
  \bibnamefont{and} \bibinfo{author}{\bibfnamefont{P.}~\bibnamefont{Borrmann}},
  \bibinfo{journal}{Phys. Rev. Lett.} \textbf{\bibinfo{volume}{88}},
  \bibinfo{pages}{053401} (\bibinfo{year}{2002}).

\bibitem[{\citenamefont{Lemmens}(1995)}]{LNAW}
\bibinfo{author}{\bibfnamefont{L.~F.} \bibnamefont{Lemmens}},
  \bibinfo{journal}{Nieuw Archief voor Wiskunde} \textbf{\bibinfo{volume}{Ser.
  4 13}}, \bibinfo{pages}{409} (\bibinfo{year}{1995}).

\bibitem[{\citenamefont{Lemmens}(1996)}]{LLPLA96}
\bibinfo{author}{\bibfnamefont{L.~F.} \bibnamefont{Lemmens}},
  \bibinfo{journal}{Phys. Lett. A} \textbf{\bibinfo{volume}{222}},
  \bibinfo{pages}{419} (\bibinfo{year}{1996}).

\bibitem[{\citenamefont{Caciuffo et~al.}(1998)\citenamefont{Caciuffo, Amoretti,
  Murani, Sessoli, Caneschi, and Gatteschi}}]{caciuffoPRL81}
\bibinfo{author}{\bibfnamefont{R.}~\bibnamefont{Caciuffo}},
  \bibinfo{author}{\bibfnamefont{G.}~\bibnamefont{Amoretti}},
  \bibinfo{author}{\bibfnamefont{A.}~\bibnamefont{Murani}},
  \bibinfo{author}{\bibfnamefont{R.}~\bibnamefont{Sessoli}},
  \bibinfo{author}{\bibfnamefont{A.}~\bibnamefont{Caneschi}}, \bibnamefont{and}
  \bibinfo{author}{\bibfnamefont{D.}~\bibnamefont{Gatteschi}},
  \bibinfo{journal}{Phys. Rev. Lett.} \textbf{\bibinfo{volume}{81}},
  \bibinfo{pages}{4744} (\bibinfo{year}{1998}).

\bibitem[{\citenamefont{Wernsdorfer et~al.}(1999)\citenamefont{Wernsdorfer,
  Ohm, Sangregorio, Sessoli, Mailly, and Paulsen}}]{wernsdorferPRL82}
\bibinfo{author}{\bibfnamefont{W.}~\bibnamefont{Wernsdorfer}},
  \bibinfo{author}{\bibfnamefont{T.}~\bibnamefont{Ohm}},
  \bibinfo{author}{\bibfnamefont{C.}~\bibnamefont{Sangregorio}},
  \bibinfo{author}{\bibfnamefont{R.}~\bibnamefont{Sessoli}},
  \bibinfo{author}{\bibfnamefont{D.}~\bibnamefont{Mailly}}, \bibnamefont{and}
  \bibinfo{author}{\bibfnamefont{C.}~\bibnamefont{Paulsen}},
  \bibinfo{journal}{Phys. Rev. Lett.} \textbf{\bibinfo{volume}{82}},
  \bibinfo{pages}{3903} (\bibinfo{year}{1999}).

\bibitem[{\citenamefont{Garanin and Chudnovsky}(1997)}]{garaninPRB56}
\bibinfo{author}{\bibfnamefont{D.~A.} \bibnamefont{Garanin}} \bibnamefont{and}
  \bibinfo{author}{\bibfnamefont{E.~M.} \bibnamefont{Chudnovsky}},
  \bibinfo{journal}{Phys. Rev. B} \textbf{\bibinfo{volume}{56}},
  \bibinfo{pages}{11102} (\bibinfo{year}{1997}).

\bibitem[{\citenamefont{Gulminelli and Chomaz}(1999)}]{PRE82_1402}
\bibinfo{author}{\bibfnamefont{F.}~\bibnamefont{Gulminelli}} \bibnamefont{and}
  \bibinfo{author}{\bibfnamefont{P.}~\bibnamefont{Chomaz}},
  \bibinfo{journal}{Phys. Rev. Lett.} \textbf{\bibinfo{volume}{82}},
  \bibinfo{pages}{1402} (\bibinfo{year}{1999}).

\bibitem[{\citenamefont{M\"{u}lken et~al.}(2001)\citenamefont{M\"{u}lken,
  Stamerjohanns, and Borrmann}}]{PRE64_047105}
\bibinfo{author}{\bibfnamefont{O.}~\bibnamefont{M\"{u}lken}},
  \bibinfo{author}{\bibfnamefont{H.}~\bibnamefont{Stamerjohanns}},
  \bibnamefont{and} \bibinfo{author}{\bibfnamefont{P.}~\bibnamefont{Borrmann}},
  \bibinfo{journal}{Phys. Rev. E} \textbf{\bibinfo{volume}{64}},
  \bibinfo{pages}{047105} (\bibinfo{year}{2001}).

\bibitem[{\citenamefont{Gulminelli and Chomaz}(2002)}]{PRE66_046108}
\bibinfo{author}{\bibfnamefont{F.}~\bibnamefont{Gulminelli}} \bibnamefont{and}
  \bibinfo{author}{\bibfnamefont{P.}~\bibnamefont{Chomaz}},
  \bibinfo{journal}{Phys. Rev. E} \textbf{\bibinfo{volume}{66}},
  \bibinfo{pages}{046108} (\bibinfo{year}{2002}).

\bibitem[{\citenamefont{Chomaz and Gulminelli}(2002)}]{PhysicaA305}
\bibinfo{author}{\bibfnamefont{P.}~\bibnamefont{Chomaz}} \bibnamefont{and}
  \bibinfo{author}{\bibfnamefont{F.}~\bibnamefont{Gulminelli}},
  \bibinfo{journal}{Physica A} \textbf{\bibinfo{volume}{305}},
  \bibinfo{pages}{330} (\bibinfo{year}{2002}).

\bibitem[{\citenamefont{Rogers and Williams}(2000)}]{williams}
\bibinfo{author}{\bibfnamefont{L.~C.~G.} \bibnamefont{Rogers}}
  \bibnamefont{and} \bibinfo{author}{\bibfnamefont{D.}~\bibnamefont{Williams}},
  \emph{\bibinfo{title}{Diffusion, Markov Processes and Martingales}}
  (\bibinfo{publisher}{Cambridge University Press}, \bibinfo{year}{2000}).

\bibitem[{\citenamefont{Bakar and Lemmens}()}]{BBLLUnPublish}
\bibinfo{author}{\bibfnamefont{B.}~\bibnamefont{Bakar}} \bibnamefont{and}
  \bibinfo{author}{\bibfnamefont{L.~F.} \bibnamefont{Lemmens}},
  \emph{\bibinfo{title}{The density of states in molecular magnets}},
  \bibinfo{note}{unpublished}.

\bibitem[{\citenamefont{Bakar and Lemmens}(2004)}]{BBLLNato2003}
\bibinfo{author}{\bibfnamefont{B.}~\bibnamefont{Bakar}} \bibnamefont{and}
  \bibinfo{author}{\bibfnamefont{L.~F.} \bibnamefont{Lemmens}},
  \bibinfo{journal}{Nato Science Series} \textbf{\bibinfo{volume}{143}},
  \bibinfo{pages}{241} (\bibinfo{year}{2004}).

\bibitem[{\citenamefont{Br\'{e}maud}(1981)}]{Bremaud}
\bibinfo{author}{\bibfnamefont{P.}~\bibnamefont{Br\'{e}maud}},
  \emph{\bibinfo{title}{Point Processes and Queues: Martingale Dynamics}}
  (\bibinfo{publisher}{Springer-Verlag, New York}, \bibinfo{year}{1981}).

\end{thebibliography}

\end{document}